\documentclass[letterpaper]{article} 
\usepackage{aaai2026}  
\usepackage{times}  
\usepackage{helvet}  
\usepackage{courier}  
\usepackage[hyphens]{url}  
\usepackage{graphicx} 
\usepackage{natbib}  
\usepackage{caption} 
\usepackage{algorithm}
\usepackage{algorithmic}
\usepackage{amsmath}
\usepackage{amsfonts}
\usepackage{makecell,multirow,diagbox}
\usepackage{tabularx}
\usepackage{newfloat}
\usepackage{listings}
\DeclareCaptionStyle{ruled}{labelfont=normalfont,labelsep=colon,strut=off} 
\floatstyle{ruled}
\newfloat{listing}{tb}{lst}{}
\floatname{listing}{Listing}
\title{Polarization Uncertainty-Guided Diffusion Model for Color Polarization Image Demosaicking}
\author{
    Chenggong Li\textsuperscript{\rm 1,\rm 2}, Yidong Luo\textsuperscript{\rm 3,\rm 4}, Junchao  Zhang\textsuperscript{\rm 1,\rm 2}\footnotemark[1], Degui Yang\textsuperscript{\rm 1,\rm 2}\thanks{Corresponding Author.}
}

\affiliations{
    \textsuperscript{\rm 1}School of Automation, Central South University, Changsha, China\\
    \textsuperscript{\rm 2}Hunan Provincial Key Laboratory of Optic-Electronic Intelligent Measurement and Control, Changsha, China\\
    \textsuperscript{\rm 3}Zhejiang University, Hangzhou, China\\
    \textsuperscript{\rm 4}School of Engineering, Westlake University, Hangzhou, China\\

\{244603040, junchaozhang, degui.yang\}@csu.edu.cn, luoyidong@westlake.edu.cn
}

\usepackage{bibentry}

\begin{document}

\maketitle

\begin{abstract}
Color polarization demosaicking (CPDM) aims to reconstruct full-resolution polarization images of four directions from the color-polarization filter array (CPFA) raw image. Due to the challenge of predicting numerous missing pixels and the scarcity of high-quality training data, existing network-based methods, despite effectively recovering scene intensity information, still exhibit significant errors in reconstructing polarization characteristics (degree of polarization, DOP, and angle of polarization, AOP). To address this problem, we introduce the image diffusion prior from text-to-image (T2I) models to overcome the performance bottleneck of network-based methods, with the additional diffusion prior compensating for limited representational capacity caused by restricted data distribution. To effectively leverage the diffusion prior, we explicitly model the polarization uncertainty during reconstruction and use uncertainty to guide the diffusion model in recovering high error regions. Extensive experiments demonstrate that the proposed method accurately recovers scene polarization characteristics with both high fidelity and strong visual perception.
\end{abstract}
\begin{links}
    \link{Code}{https://github.com/JJGNB/PUGDiff}
\end{links}
\section{Introduction}
Polarization imaging can reveal unique object properties, such as material and reflectivity. By fully leveraging the degree of polarization (DOP) and angle of polarization (AOP), polarization imaging has been increasingly applied in fields like object detection, reflectivity removal, and 3D reconstruction \cite{luo2025cpifuse, yao2025polarfree, wu2025glossy}. In practical applications, the photographer captures a mosaic array with polarization information using a division-of-focal-plane (DOFP) camera \cite{rebhan2019principle}. Fig. 1 shows the operation of the DOFP camera, which samples pixels from four polarization directions (${0^\circ }$, ${45^\circ }$, ${90^\circ }$, ${135^\circ }$) into a mosaic array following an RGGB order, and the sampling process can be expressed as:
\begin{equation}
	\begin{aligned}
		y = Ax+n ,
	\end{aligned}
	\label{eq:1}
\end{equation}
 where $A$ denotes the sampling operator, $x \in {\mathbb{R}^{12\times H \times W }}$, $y \in {\mathbb{R}^{1\times H \times W}}$ and $n \in {\mathbb{R}^{1\times  H \times W }}$ represent the ground truth full-resolution color image, the mosaic array and the sensor noise, respectively. To obtain a full-resolution color image for better analysis of polarization characteristics, we need to recover the pixels lost during the sampling from the mosaic array, i.e., color polarization demosaicking (CPDM). 
\begin{figure}[!t]
	\begin{center}
		\includegraphics[width=\linewidth]{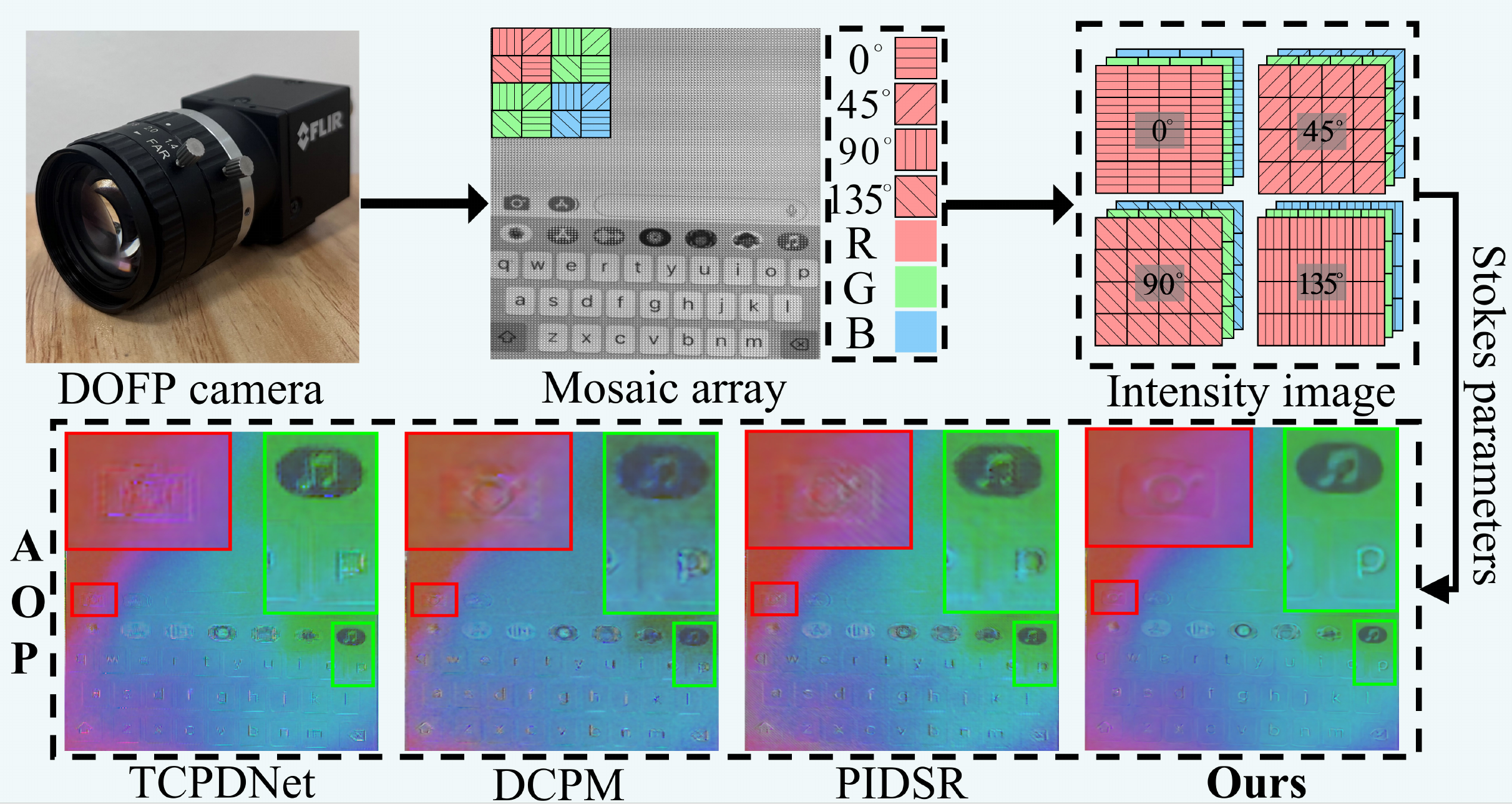}
	\end{center}
	\caption{The workflow of color polarization imaging. The second row presents the AOP results of different CPDM methods, where AOP reflects the accuracy of polarization property reconstruction. Our method exhibits the least noise interference and achieves the most visually faithful results.}
	\label{fig:1}
\end{figure}

CPDM is a challenging inverse problem in polarization imaging. Some prior-based interpolation algorithms \cite{morimatsu2020monochrome, wu2021polarization} and model-based optimization algorithms \cite{wen2021sparse, luo2024learning} have attempted to address this issue, but the reconstruction accuracy remains limited. Using neural networks ${f_\theta }$  to address CPDM is a promising approach \cite{zhang2018learning, sun2021color} and the demosaicking process can be formulated as:
\begin{equation}
	\begin{aligned}
		\begin{array}{l}
			\tilde x = I(y)\\
			x = {f_\theta }(\tilde x) + \varepsilon \eta 
		\end{array}, 
	\end{aligned}
	\label{eq:2}
\end{equation}
where $I$ denotes the initial interpolation that restores the mosaic array into four direction intensity images $\tilde x \in {\mathbb{R}^{12\times H \times W }}$, $\varepsilon$ follows the normal distribution with zero-mean and unit-variance, and $\eta$ denotes the demosaicking error, which we term the uncertainty in the demosaicking process. By employing a predefined sampling operator to generate simulated datasets, network-based methods can learn to recover missing pixels from the data. However, even though these methods can reconstruct the intensity image well, they exhibit significant errors in recovering polarization characteristics like AOP, as shown in Fig. \ref{fig:1}. This is because current network-based methods learn demosaicking capabilities from limited datasets \cite{morimatsu2020monochrome, qiu2021linear, li2025demosaicking}, which lack scene diversity and adequate scale, making the learned data priors insufficient to support these methods in achieving superior performance.

\begin{figure*}[!t]
	\begin{center}
		\includegraphics[width=\linewidth]{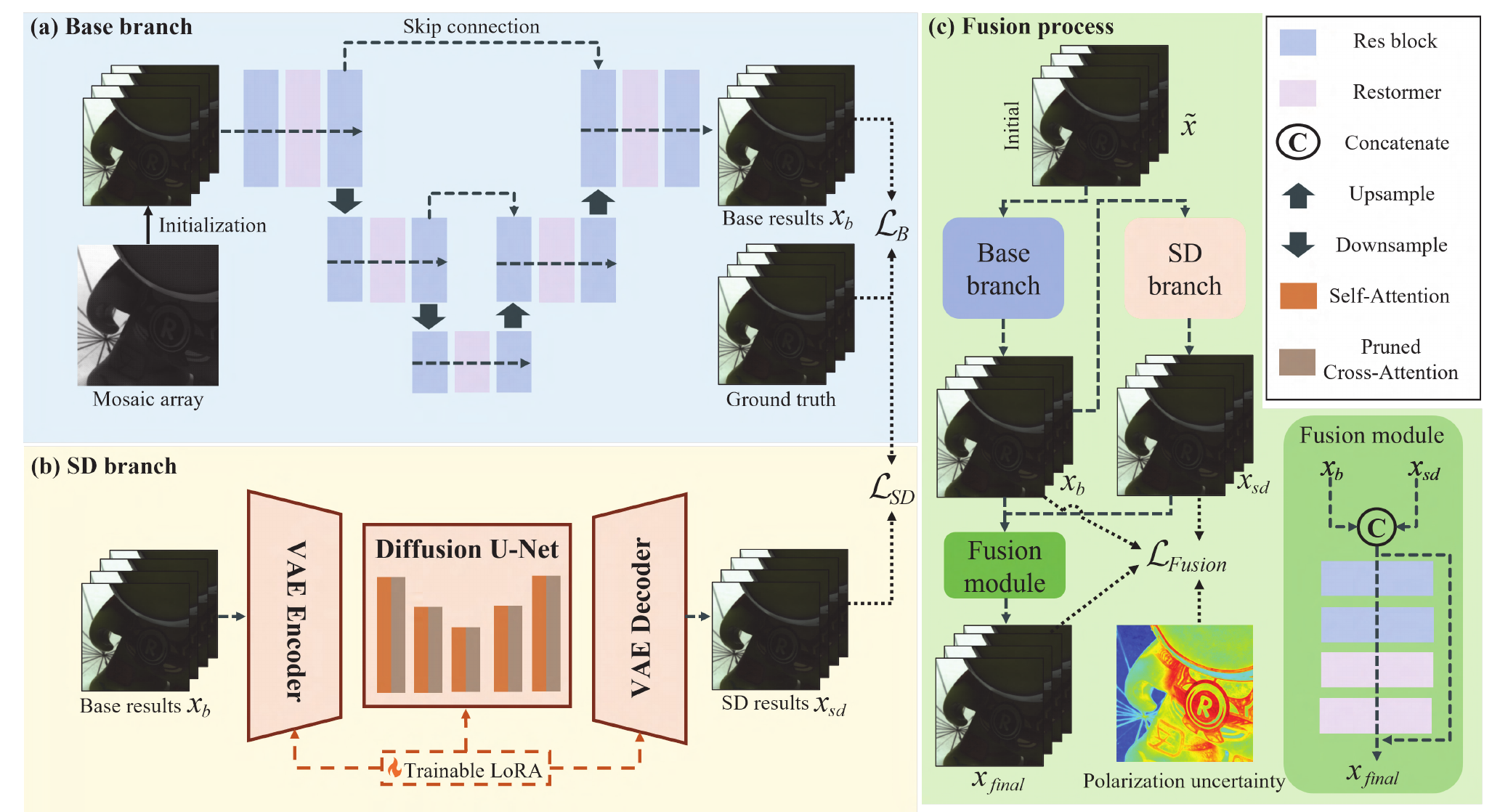}
	\end{center}
	\caption{The framework of the proposed PUGDiff. (a) The architecture of the base branch, the mosaic array is initialized as a low-quality color polarization image and fed into the base branch for processing, and the network is updated via $\mathcal{L}_B$; (b) The architecture of the SD branch, results of the base branch is further refined by the SD branch. LoRA layers are injected into SD to adapt it to the CPDM task, while the cross-attention modules are pruned to improve efficiency, the LoRA parameters are updated via $\mathcal{L}_{SD}$; (c) The fusion process of the two branches, which is also used during inference. The contribution of each branch is adaptively weighted by polarization uncertainty. In low-uncertainty regions, the base branch is favored for high fidelity; in high-uncertainty regions, the SD branch is used to enhance the polarization property.}
	\label{fig:2}
\end{figure*}

To achieve better demosaicking performance, we propose a polarization uncertainty-guided diffusion model, namely PUGDiff. PUGDiff is a dual-branch network shown in Fig. \ref{fig:2}, a base branch trained from scratch provides the model with fundamental demosaicking capabilities, while the other branch consists of a text-to-image model (i.e.,  Stable Diffusion, SD), called SD branch. The SD branch is trained using Low-Rank Adaptation (LoRA) \cite{hu2022lora} to maximally preserve its diffusion prior, which helps overcome the limitations of existing data priors. We explicitly model a polarization uncertainty estimator to effectively leverage the two branches. Specifically, in regions with low uncertainty, the base branch can produce high-fidelity results, while in regions with high uncertainty, the SD branch helps correct polarization errors. The two branches are fused by a fusion module to generate the final output, which is trained using an uncertainty-guided loss, the loss transfers the uncertainty map into weights for different loss terms and thus directs the model to flexibly allocate the contributions of the two branches to the final result based on polarization uncertainty. 

We summarize the primary contributions as follows:
\begin{itemize}
    \item We introduce the text-to-image model into the CPDM, leveraging its diffusion prior learned from large-scale natural images to break through the limitations of the original data distribution.
    \item We model the uncertainty in the demosaicking process based on polarization characteristics, and transform it into an uncertainty-guided loss to supervise network fusion, enabling the network to adaptively select the dominant branch based on the polarization uncertainty.
    \item Qualitative and quantitative experiments on both simulated and real-world images demonstrate that our method achieves state-of-the-art (SOTA) performance.
\end{itemize}

\section{Related Work}
\textbf{Color Polarization Demosaicking.} Early research efforts on CPDM primarily relied on interpolation algorithms \cite{li2019demosaicking, morimatsu2020monochrome} or dictionary learning \cite{wen2021sparse, luo2023sparse, luo2024learning}. With the blooming of deep learning, researchers have gradually shifted their focus toward network-based approaches. Nguyen et al. propose a two-stage demosaicking convolutional neural network (CNN) \cite{nguyen2022two}, achieving joint demosaicking for color and polarization.  Guo et al. design an attention-based progressive discrimination generative adversarial network (GAN) to alleviate artifacts during the demosaicking process \cite{guo2024attention}. Zhou et al. incorporate polarization characteristics into the network design, achieving mutual enhancement between demosaicking and polarization super-resolution through a carefully designed pipeline \cite{zhou2025pidsr}. Li et al. propose a demosaicking customized diffusion model, which achieves more robust results through distribution modelling \cite{li2025demosaicking}. However, existing network-based methods only learn demosaicking capabilities from small-scale simulated datasets, leading to a performance bottleneck that makes them struggle to handle more complex and diverse scenarios.

\noindent \textbf{Diffusion Prior in Inverse Problems.} Using diffusion priors to assist in solving inverse problems is currently a hot topic in low-level vision. Many diffusion paradigms have been successfully applied to tasks like denoising, deblurring, inpainting, compressed sensing, super-resolution, etc \cite{kawar2022denoising, sun2023improving, wu2024seesr, chihaoui2024blind, sun2025pixel, chen2025invertible, chen2025adversarial}. Chung et al. integrate inverse problem solving into the diffusion solver through approximation of the posterior sampling, enabling effective handling of noisy inverse problems \cite{chung2022diffusion}. Wang et al. utilize a pre-trained diffusion model to progressively restore the null-space of images while ensuring data consistency by range-space constraints \cite{wang2022zero}. Wu et al. introduce LoRA into Stable Diffusion and utilize variational score distillation loss to align the diffusion model with natural image priors \cite{wu2024one}. Yue et al. design a deep noise predictor based on diffusion inversion, enabling noise customization for super-resolution and achieving high perceptual quality results \cite{yue2025arbitrary}. These achievements motivate us to explore the application of diffusion priors to polarization demosaicking, aiming to overcome the performance bottleneck caused by insufficient datasets.

\section{Methodology}
In this paper, we propose a dual-branch diffusion model guided by polarization uncertainty. As shown in Fig. \ref{fig:2}, we first train a demosaicking network $f_b$ from scratch, using a CNN-Transformer \cite{zamir2022restormer} hybrid U-Net architecture. $f_b$ serves as the algorithm's base branch, aiming to meet the high fidelity requirements of the CPDM task. The process of the base branch can be expressed as:
\begin{equation}
	\begin{aligned}
		x_b = {f_b}(\tilde x),
	\end{aligned}
	\label{eq:3}
\end{equation}
$x_b \in {\mathbb{R}^{12\times H \times W  }}$ is then fed into a modified SD (SD branch) to leverage its inherent diffusion prior for further refinement, the above process can be expressed as:
\begin{equation}
	\begin{aligned}
		x_{sd} = {f_{sd}}(x_b). 
	\end{aligned}
	\label{eq:4}
\end{equation}

To effectively utilize the two branches for better demosaicking performance, we develop a polarization uncertainty model and use the uncertainty to guide the fusion of the two branches' outputs. The fusion process can be formulated as:
\begin{equation}
	\begin{aligned}
		x_{final} = {f_{fuse}}(x_b,x_{sd}). 
	\end{aligned}
	\label{eq:5}
\end{equation}

In the following, we detail the modifications made to the SD and the derivation of the polarization uncertainty model. The loss functions for each module are then explained.
\subsection{Modified SD}
The SD consists of a variational autoencoder (VAE) and a latent diffusion model (LDM) shown in Fig. \ref{fig:2} (b). Since SD processes natural images, we treat the four direction intensity images as four separate batches and feed them into the SD model sequentially. The data flow of SD can be expressed as:
 \begin{equation}
\begin{array}{l}
{z_b}{\rm{ }} = E({x_b}),\\
{z_{sd}} = LDM({z_b}),\\
{x_{sd}} = D({z_{sd}}),
\end{array}
 	\label{eq:6}
 \end{equation}
 where $E$ and $D$ denote the VAE encoder and decoder, respectively, with $z_b$ and $z_{sd}$ representing the image features in the latent space. The LDM obtains the final result using only one step, which can be expressed as:
  \begin{equation}
 	\begin{aligned}
{z_{sd}} = LDM({z_b}) \buildrel \Delta \over = \frac{{{z_b} - {\beta _T}\psi  ({z_b},T)}}{{{\alpha _T}}},
 	\end{aligned}
 	\label{eq:7}
 \end{equation}
where $T$ is the time, $\alpha _T$ and $\beta _T$ denotes the noise scheduler of the SD, and $\psi$ represents the diffusion U-Net. In the context of CPDM, text prompts are unnecessary for the SD, hence, we remove the text encoder and cross-attention module in the denoising U-Net, improving efficiency without compromising demosaicking performance. 
\begin{figure*}[!t]
	\begin{center}
		\includegraphics[width=\linewidth]{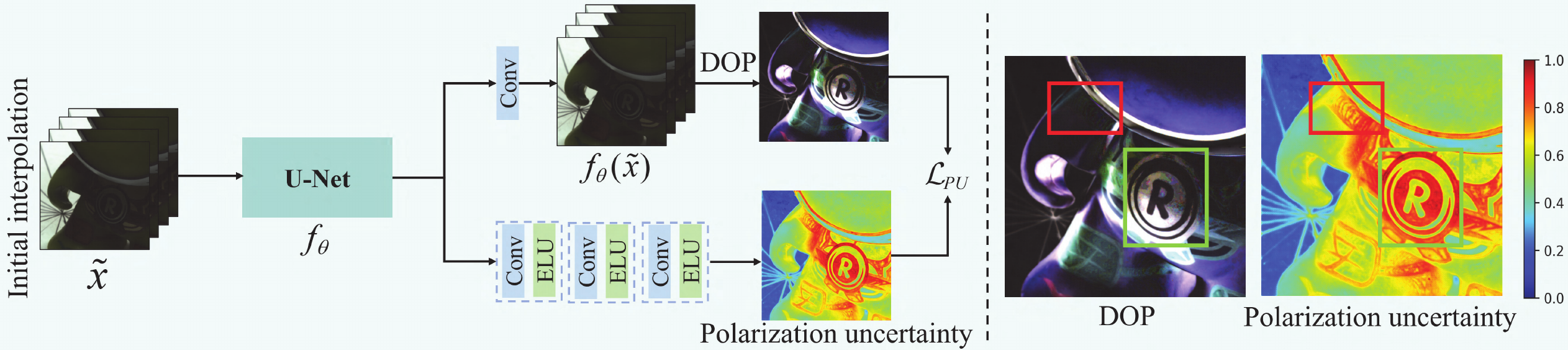}
	\end{center}
	\caption{The architecture of the uncertainty estimation network. The network's backbone shares the same architecture as the base branch, but augmented with an additional estimation head to output polarization uncertainty. The entire network is updated via $\mathcal{L}_{PU}$, and during the training phase in Fig. \ref{fig:2} (c), only the polarization uncertainty is used. The right part of the figure shows the reconstructed DOP and the polarization uncertainty. High values in the heatmap indicate regions where DOP reconstruction is inaccurate.}
	\label{fig:3}
\end{figure*}

Instead of training from scratch, we inject trainable LoRA modules into the VAE and diffusion U-Net, and fine-tune them on simulated datasets to accommodate the CPDM task. Training with LoRA effectively preserves the diffusion prior of the SD, enabling the network to leverage additional knowledge to recover lost pixels. This helps overcome the performance bottleneck caused by limited CPDM datasets and further enhances the models' generalization capability. 

\subsection{Polarization Uncertainty Model}
Although SD can produce high-quality perceptual results, it compresses the image, leading to information loss and failing to meet the high-fidelity requirements of CPDM. In contrast, while the base branch recovers four polarization directions well, the polarization properties (DOP and AOP) are derived through nonlinear computation, which amplifies reconstruction errors. This finding inspires us to design a rule for fusing the outputs of the two branches. For regions with small reconstruction errors from the base branch, we prefer its results to ensure fidelity; for regions with large errors, we place more confidence in the SD output to improve the visual faithfulness of polarization properties.

Therefore, from the perspective of polarization reconstruction error (uncertainty), we establish a polarization uncertainty model to explicitly characterize the uncertainty. We reformulate Eq. (\ref{eq:2}) as:
\begin{equation}
	\begin{aligned}
		{x_i} = {f_\theta }({\tilde x_i}) + \varepsilon \eta, i = {0^\circ }, {45^\circ }, {90^\circ }, {135^\circ },
	\end{aligned}
	\label{eq:8}
\end{equation}
where $\eta$ is the intensity uncertainty, and $\varepsilon$ follows the normal distribution with zero-mean and unit-variance, the subscript $i$ indicates the polarization images at different directions. Eq. (\ref{eq:8}) can be expressed as ${x_i} \sim N({f_\theta }({\tilde x_i}),{\rm{ }}\eta^2 )$. According to Collett's literature \cite{collett2005field}, the Stokes parameters are defined as:
\begin{equation}
	\begin{aligned}
\begin{array}{l}
	{S_0} = \frac{1}{2}({x_{{0^\circ }}} + {x_{{{45}^\circ }}} + {x_{{{90}^\circ }}} + {x_{{{135}^\circ }}})\\
	{S_1} = {x_{{0^\circ }}} - {x_{{{90}^\circ }}}\\
	{S_2} = {x_{{{45}^\circ }}} - {x_{{{135}^\circ }}}\\
	DOP = \frac{{\sqrt {{S_1}^2 + {S_2}^2} }}{{{S_0}}}\\
	AOP = \frac{1}{2}\arctan (\frac{{{S_2}}}{{{S_1}}})
\end{array}.
	\end{aligned}
	\label{eq:9}
\end{equation}

Substituting Eq. (\ref{eq:8}) into Eq. (\ref{eq:9}) and using the properties of the normal distribution gives:
\begin{equation}
	\begin{aligned}
{S_0} \sim N({\tilde S_0},{\rm{ }}\eta^2 ),{\rm{ }}{S_1} \sim N({\tilde S_1},{\rm{ }}2\eta^2 ),{\rm{ }}{S_2} \sim N({\tilde S_2},{\rm{ }}2\eta^2 ),
	\end{aligned}
	\label{eq:10}
\end{equation}
where
\begin{equation}
	\begin{aligned}
		\begin{array}{l}
{\tilde S_0} \buildrel \Delta \over = \frac{1}{2}({f_\theta }({\tilde x_{{0^\circ }}}) + {f_\theta }({\tilde x_{{{45}^\circ }}}) + {f_\theta }({\tilde x_{{{90}^\circ }}}) + {f_\theta }({\tilde x_{{{135}^\circ }}}))\\
{\rm{ }}{\tilde S_1} \buildrel \Delta \over = {f_\theta }({\tilde x_{{0^\circ }}}) - {f_\theta }({\tilde x_{{{90}^\circ }}}),{\rm{ }}{\tilde S_2} \buildrel \Delta \over = {f_\theta }({\tilde x_{{{45}^\circ }}}) - {f_\theta }({\tilde x_{{{135}^\circ }}})
\end{array}.
	\end{aligned}
	\label{eq:11}
\end{equation}

Since $f_\theta$ has already reconstructed the intensity image ($S_0$) effectively, we can assume that the estimate of $S_0$ is accurate \footnote{This observation is also supported by the visual comparisons in the experiment section, where most CPDM methods achieve $S_0$ results close to the ground truth.}, and thus use $\tilde S_0$ as an approximation of the ground truth $S_0$, i.e. $\tilde S_0\buildrel\Delta \over =S_0$. Subsequently, we can derive:
\begin{equation}
	\begin{aligned}
		\begin{array}{l}
			\frac{{{S_1}}}{{{S_0}}} \sim N(\frac{{{{\tilde S}_1}}}{{{{\tilde S}_0}}},{\rm{ }}\frac{{2\eta^2 }}{{{{\tilde S}^2_0}}}),{\rm{ }}\frac{{{S_2}}}{{{S_0}}} \sim N(\frac{{{{\tilde S}_2}}}{{{{\tilde S}_0}}},{\rm{ }}\frac{{2\eta^2 }}{{{{\tilde S}^2_0}}}).
		\end{array}
	\end{aligned}
	\label{eq:12}
\end{equation}

Based on Eqs. (\ref{eq:9}) and (\ref{eq:12}), the DOP (denoted as $\phi $) follows a Rice distribution of the form \cite{Hwang_2025_ICCV}:
\begin{equation}
	\begin{aligned}
		\begin{array}{l}
\phi  \sim Rice(\tilde \phi ,{\rm{ }}\frac{{\sqrt 2 \eta }}{{{S_0}}}),
		\end{array}
	\end{aligned}
	\label{eq:13}
\end{equation}
and we denote $\frac{{\sqrt 2 \eta }}{{{S_0}}}$ as ${\eta _p}$, which represents the polarization uncertainty we aim to model in this paper. Compared to the intensity uncertainty $\eta$, ${\eta _p}$ provides more direct polarization information for the subsequent fusion process.

The probability density function (PDF) of Eq. (\ref{eq:13}) can be expressed as:
\begin{equation}
	\begin{aligned}
		\begin{array}{l}
			P(\phi |\tilde \phi ,{\rm{ }}{\eta _p}) = \frac{\phi }{{{\eta _p}^2}}\exp ( - \frac{{({\phi ^2} + {{\tilde \phi }^2})}}{{2{\eta _p}^2}}){I_0}(\frac{{\phi \tilde \phi }}{{{\eta _p}^2}}),
		\end{array}
	\end{aligned}
	\label{eq:14}
\end{equation}
where $I_0$ denotes the modified Bessel function of the first kind, order zero. In applications, the signal amplitude is much larger than the uncertainty; therefore, according to the definition of the modified Bessel function \cite{bowman2012introduction}, $I_0$ has an asymptotic form for large arguments, and Eq. (\ref{eq:14}) can be approximated as \footnote{The detailed derivation can be found in the supplementary material.} :
\begin{equation}
	\begin{aligned}
		\begin{array}{l}
P(\phi |\tilde \phi ,{\rm{ }}{\eta _p}) \approx \sqrt {\frac{\phi }{{2\pi \tilde \phi {\eta _p}^2}}} \exp ( - \frac{{{{(\phi  - \tilde \phi )}^2}}}{{2{\eta _p}^2}}).
		\end{array}
	\end{aligned}
	\label{eq:15}
\end{equation}

To obtain $\eta _p$, we use an uncertainty estimation network to directly predict it through supervised learning shown in Fig. \ref{fig:3}. The estimation network's weights are initialized from the base branch for consistency, and the optimization objective is to minimize the negative log likelihood of Eq. (\ref{eq:15})\footnote{The detailed derivation can be found in the supplementary material.}, i.e.,
\begin{equation}
	\begin{aligned}
		\begin{array}{l}
		{{\mathcal L}_{PU}} = \frac{1}{2}(\ln \tilde \phi  - \ln \phi ) + 2s + \frac{1}{2}\exp ( - 2s)\left\| {\tilde \phi  - \phi } \right\|_2^2,
		\end{array}
	\end{aligned}
	\label{eq:16}
\end{equation}
where $s$ denotes the log polarization uncertainty $s = \ln {\eta _p}$.
\begin{table*}[!t]
	\centering
	\small
	\begin{tabular}{c|llllllll}
		\hline
		&Method & $PSNR_{mean}$  & $PSNR_{S_{0}}$  & $PSNR_{DOP}$ & $MAE$  & $SSIM_{mean}$  & $SSIM_{S_{0}}$  & $SSIM_{DOP}$ \\
		\hline
		\multirow{7}{*}{MQ's}&
		Polanalyser &38.2149	&35.2560	&32.5532 &14.8220 &0.9653	&0.9541	&0.8387\\
		&NLCSR         &36.8877	&33.5610	&34.0964 &11.1297	&0.9523	&0.9310	&0.8557\\
		&CPDNet     &39.4374	&36.6062	&32.6586 &12.5138	&0.9700	&0.9631	&0.8465\\
		&TCPDNet  &42.5602	&39.7694	&36.0613 &10.1765	&0.9815	&0.9765	&0.8921\\
		&DCPM    &42.0221	&40.0263	&36.1496 &11.7376	&0.9807	&0.9759	&0.8958\\
		&PIDSR      &42.4990	&\textbf{40.6269}	&36.5413 &9.6463	&0.9822	&\textbf{0.9794}	&0.8950\\
		&Ours      &\textbf{42.6794}	&39.4785	&\textbf{37.4700} &\textbf{9.2029}	&\textbf{0.9827}	&0.9768	&\textbf{0.9059}\\
		\hline
		\multirow{7}{*}{PIDSR's}&
		Polanalyser &41.0285	&39.4167	&35.8349 &17.9892	&0.9655	&0.9621	&0.8347\\
		&NLCSR         &40.0285	&38.5526	&36.0323 &16.0549	&0.9601	&0.9569	&0.8775\\
		&CPDNet     &41.4389	&39.6154	&37.0999 &\textbf{14.9234}	&0.9694	&0.9639	&0.8529\\
		&TCPDNet  &43.1434	&41.2190	&39.1320 &15.5998	&0.9785	&0.9734	&0.8883\\
		&DCPM    &43.4987	&41.7152	&38.4340 &15.5325	&0.9804	&0.9750	&0.8791\\
		&PIDSR      &43.0540	&41.2950	&39.8428 &16.4804	&0.9792	&0.9741	&0.9015\\
		&Ours      &\textbf{44.2325}	&\textbf{41.9823}	&\textbf{40.6696} &17.0324	&\textbf{0.9815}	&\textbf{0.9753}	&\textbf{0.9107}\\
		\hline
		\multirow{7}{*}{DCPM's}&
		Polanalyser &36.6754	&35.4464	&29.3446 &17.2526	&0.9653	&0.9652	&0.7344\\
		&NLCSR         &37.8591	&36.8419	&29.5389 &15.6778	&0.9688	&0.9693	&0.7294\\
		&CPDNet     &39.3842	&39.4640	&29.9120 &16.2399	&0.9767	&0.9806	&0.7463\\
		&TCPDNet  &40.2257	&40.7204	&30.5104 &15.9773	&0.9789	&0.9829	&0.7532\\
		&DCPM    &43.2512	&\textbf{43.7558}	&32.5309 &14.7762	&0.9857	&0.9870	&0.7765\\
		&PIDSR      &40.5454	&42.5008	&31.7020 &14.5157	&0.9815	&0.9856	&0.7624\\
		&Ours      &\textbf{43.3532}	&42.8488	&\textbf{33.2396} &\textbf{12.6792} &\textbf{0.9871}	&\textbf{0.9877}	&\textbf{0.7934}\\
		\hline
		
	\end{tabular}
	\caption{Quantitative comparison with state-of-the-art methods. The higher value means better effects. The bold black indicates the best result. The proposed method shows the overall optimal performance.}
	\label{tab:1}
\end{table*}
\subsection{Uncertainty Guided Loss Function}
We use the mean square error (MSE) loss to train both the base branch $f_b$ and SD branch $f_{sd}$, i.e.,
\begin{equation}
	\begin{aligned}
		\begin{array}{l}
{{\cal L}_{B\& SD}} = \left\| {x - {x_r}} \right\|_2^2,{\rm{ }}{x_r} \in \{ {x_b},{\rm{ }}{x_{sd}}\} ,
		\end{array}
	\end{aligned}
	\label{eq:17}
\end{equation}
where $x \in {\mathbb{R}^{ 12 \times H \times W }}$ denotes the ground truth of four directions.

To better leverage the strengths of both branches, we define the following uncertainty-guided fusion loss shown in Fig. \ref{fig:2} (c):
\begin{equation}
	\begin{aligned}
		\begin{array}{l}
{{\cal L}_{Fusion}} = \bar s\left\| {{x_{final}} - {x_{sd}}} \right\|_2^2 + (1 - \bar s)\left\| {{x_{final}} - {x_b}} \right\|_2^2,
		\end{array}
	\end{aligned}
	\label{eq:18}
\end{equation}
where $\bar s$ is the normalized $s$. The above function fuses the two branches based on polarization uncertainty. In regions with small ${\eta _p}$, the base branch ensures high fidelity, while in regions with large polarization estimation uncertainty, the SD branch is emphasized to improve the quality of polarization reconstruction. ${{\cal L}_{Fusion}}$ implicitly incorporates polarization uncertainty as a gating mechanism within the fusion module. During inference, there is no need to explicitly output the uncertainty, thereby reducing the network's forward computation.
\begin{figure*}[!t]
	\begin{center}
		\includegraphics[width=\linewidth]{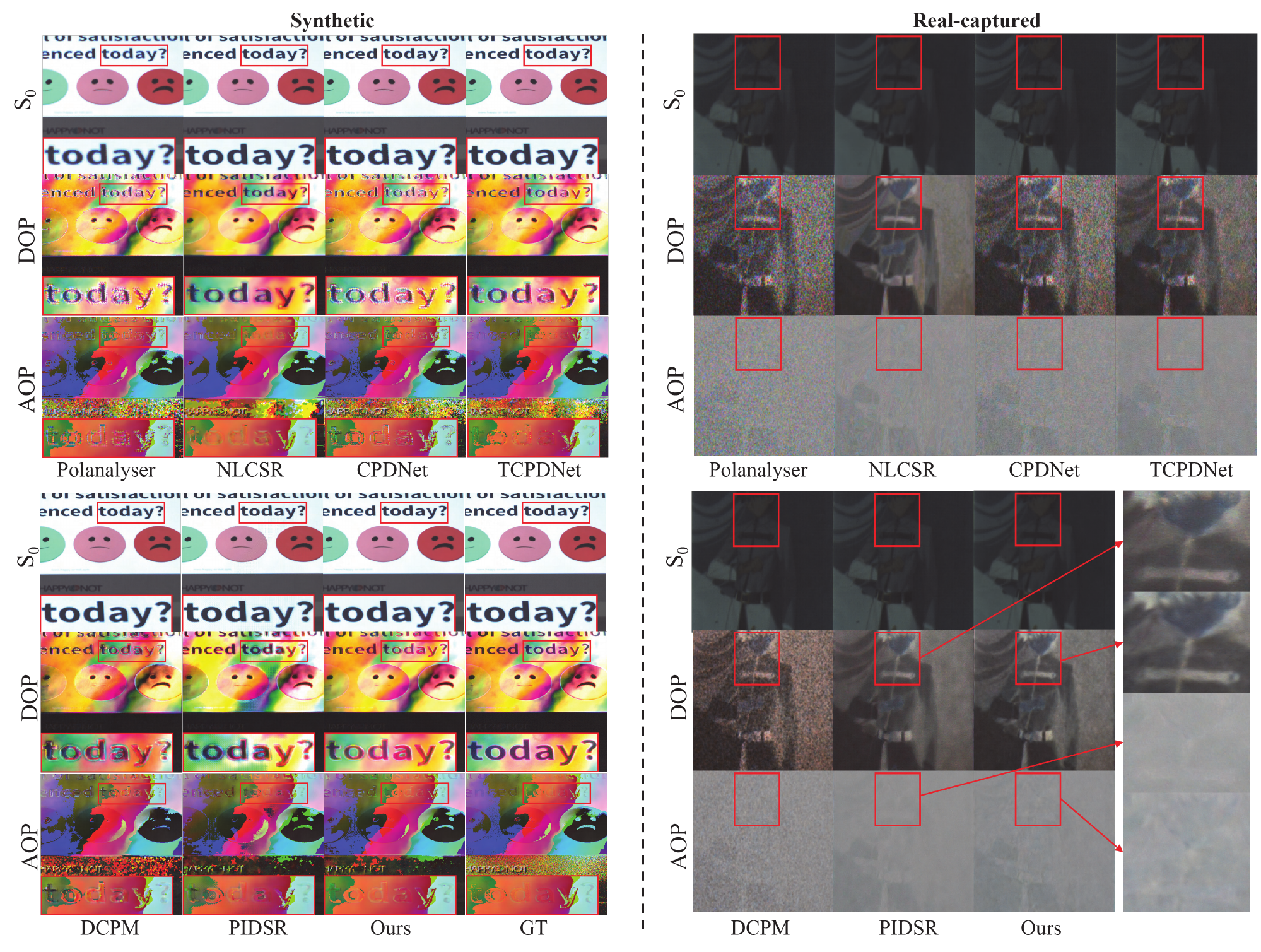}
	\end{center}
	\caption{Visual comparisons for CPDM of different methods. Both on synthetic and real-captured images, our method achieves the most outstanding polarization reconstruction performance in terms of AOP and DOP.}
	\label{fig:4}
\end{figure*}
\begin{figure}[!t]
	\begin{center}
		\includegraphics[width=\linewidth]{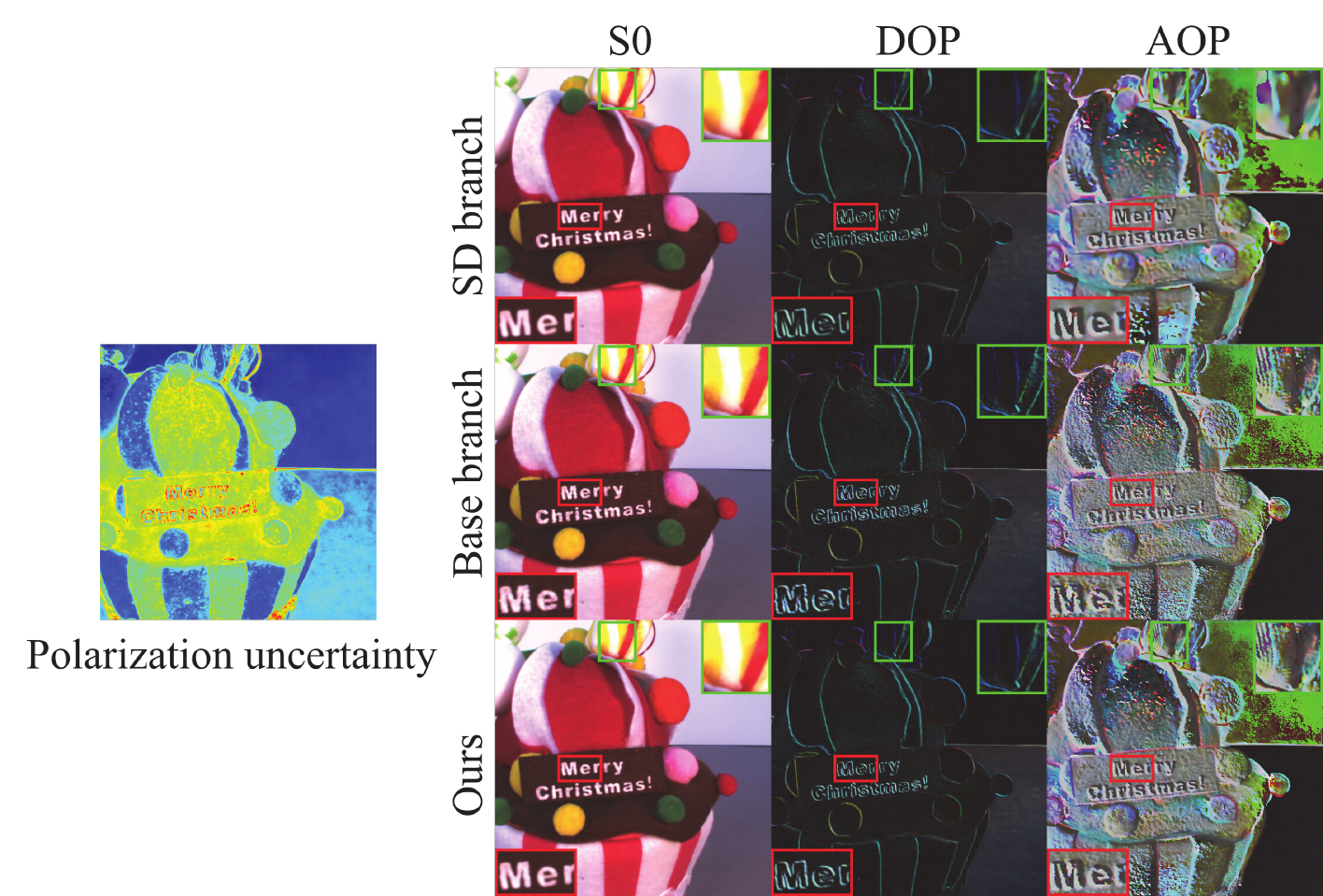}
	\end{center}
	\caption{Visualization of results from different branches. Our method fuses the two branches to achieve complementary strengths: using the SD branch in high-uncertainty regions to improve reconstruction, and the base branch in low-uncertainty regions to preserve fidelity.}
	\label{fig:5}
\end{figure}

\section{Experiments}
\subsection{Implementation details}
For the experiment part, the proposed PUGDiff is trained with Monno's \cite{morimatsu2020monochrome}, Qiu's \cite{qiu2021linear} and PIDSR \cite{zhou2025pidsr} synthetic datasets. Cropping, flipping and rotation are applied to augment the training data. The final training set consists of 28020 image patches of size $256  \times 256$. The mosaic arrays are first interpolated using Polanalyser \cite{maeda2019polanalyser} for initial reconstruction before being fed into the network. To ensure training stability, the base branch, SD branch, and uncertainty estimation network are trained separately. For the rank of LoRA in the SD branch, we set it to 4. After training the above modules, we proceed to train the fusion module. All models are trained for 200k iterations with a fixed learning rate of $5 \times {10^{ - 5}}$ and a batch size of 2. All experiments are conducted on an NVIDIA RTX 4090 GPU.
\subsection{Experimental configurations}
We conduct tests on the Monno's, Qiu's (MQ), PIDSR, and DCPM \cite{li2025demosaicking} simulated datasets, and validate the generalization on the real-captured mosaic array. The proposed method is compared with Polanalyser \cite{maeda2019polanalyser}, NLCSR \cite{luo2024learning}, CPDNet \cite{wen2019convolutional}, TCPDNet \cite{nguyen2022two}, DCPM \cite{li2025demosaicking} and PIDSR \cite{zhou2025pidsr}. We evaluate the quality of CPDM with PSNR and SSIM. The subscript ``mean'' indicates that the metric is calculated on the intensity images of four directions and then averaged, while ``$S_0$'', ``DOP'' and ``AOP'' denote the metrics calculated for $S_0$, DOP and AOP, respectively. The higher value of both metrics means better CPDM effects. We also use Mean Angular Error (MAE) to evaluate AOP quality, where lower values indicate better performance.
\subsection{Comparison with State-of-the-art Methods}
The CPDM results of all methods are shown in Fig. \ref{fig:4}. Although these methods successfully reconstruct the intensity images ($S_0$), the AOP and DOP, which reflect the scene's polarization properties, still exhibit significant errors. On the simulated data, our method produces the sharpest text edges and exhibits the smallest deviation from the ground truth. On the real-captured image, most algorithms are overwhelmed by noise. The closest performer is PIDSR; upon magnification, our method shows clearer DOP details on clothing and facial regions, and more distinct AOP compared to PIDSR. Table \ref{tab:1} presents the quantitative results on three datasets, where our method achieves the best overall performance, particularly in DOP and AOP metrics. This improvement is attributed to the polarization uncertainty-guided diffusion prior, which enables our method to overcome the performance ceiling of conventional network-based methods.
\subsection{Ablation Studies}
\textbf{Importance of polarization uncertainty in the dual-branch framework.} The network adjusts the contribution of the base branch and the SD branch to the final result under the guidance of polarization uncertainty. In Fig. \ref{fig:5}, we visualize the outputs of both branches along with the corresponding polarization uncertainty maps. The polarization uncertainty in the text regions is relatively high, which is also reflected in the AOP and DOP results of the base branch. Consequently, the network assigns greater weight of the SD branch in these areas. In contrast, for regions with low uncertainty (as highlighted in the green box), the base branch is capable of accurately reconstructing the polarization characteristics. In such areas, the SD branch tends to produce over-smoothed results, degrading demosaicking accuracy. Therefore, the network gives higher confidence to the base branch, enhancing overall fidelity.
\begin{table}[!t]
	\centering
	
		\small
	\begin{tabular}{lllll}
		\hline
		&Method & $PSNR_{S_{0}}$  & $PSNR_{DOP}$  & $PSNR_{AOP}$  \\
		\hline
		\multirow{3}{*}&
		Intensity &\textbf{42.1747} &40.2787 &15.8751\\
		&Intensity ($S_0$)         &42.1165 &40.3096 &16.4514\\
		&Polarization   &41.9823	&\textbf{40.6696}	&\textbf{16.8207}\\
		\hline
		
	\end{tabular}
	\caption{Ablation study on the forms of uncertainty. ``Intensity'', ``Intensity ($S_0$)'' and ``Polarization'' denote the uncertainty computed from the four direction polarization images, $S_0$, and DOP, respectively. The bold black indicates the best result.}
	\label{tab:2}
\end{table}
\begin{table}[!t]
	\centering
	
		\small
	\begin{tabular}{lllll}
		\hline
		&Method & $PSNR_{S_{0}}$  & $PSNR_{DOP}$  & $PSNR_{AOP}$  \\
		\hline
		\multirow{5}{*}&
		All weight &--	&--	&--	\\
		&Only U-Net         &40.8976	&40.2579	&16.0951 \\
		&rank 2    &--	&--	&--	\\
		&rank 4 (Ours)    &41.9823	&\textbf{40.6696}	&\textbf{16.8207} \\
		&rank 8    &\textbf{42.0145}	&40.6652	&16.5832\\
		\hline
		
	\end{tabular}
	\caption{Ablation study on the SD configuration. ``All weight'' denotes full-parameter fine-tuning, ``Only U-Net'' indicates that only the diffusion U-Net is updated, ``rank'' is a hyperparameter of LoRA and ``--'' indicates training failure.  The bold black indicates the best result.}
	\label{tab:3}
\end{table}

\begin{figure}[!t]
	\begin{center}
		\includegraphics[width=\linewidth]{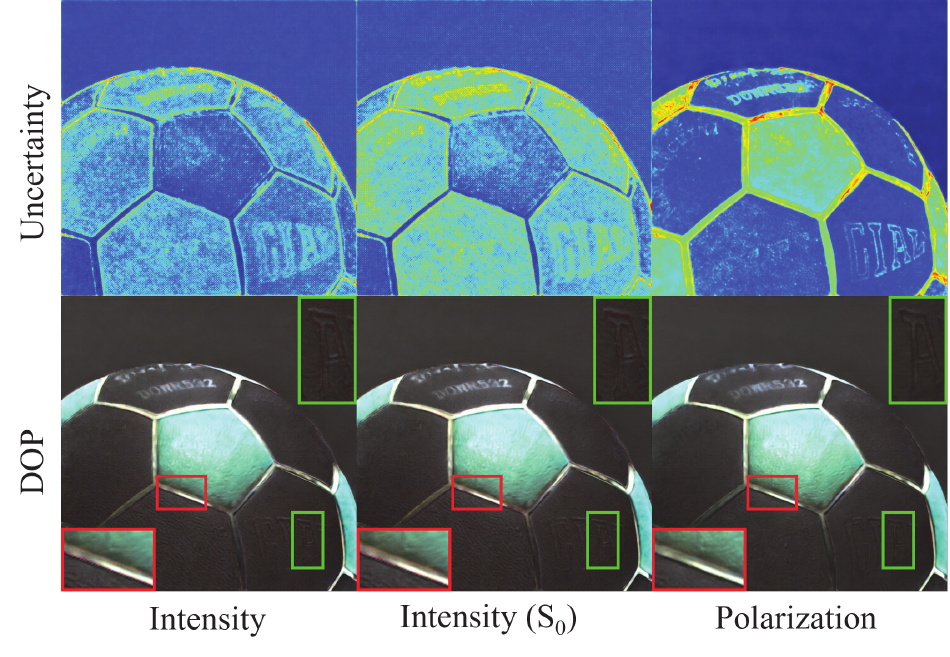}
	\end{center}
	\caption{Visualization of different forms of uncertainty. Polarization uncertainty provides more direct information about polarization reconstruction, thus leading to the most significant improvement in DOP.}
	\label{fig:6}
\end{figure}

\noindent \textbf{Superiority of modelling uncertainty from the polarization perspective.} In the main text, we explicitly model polarization uncertainty based on the PDF of DOP. To demonstrate its effectiveness in recovering polarization information, we also derive uncertainty for the four polarization directions and for $S_0$ using Eqs. (\ref{eq:8}) and (\ref{eq:10})\footnote{The derivation can be found in the supplementary material.}, respectively. The uncertainty maps are shown in Fig. \ref{fig:6}. The uncertainty derived from intensity images reflects errors in reconstructing intensity information. However, in the context of CPDM, our primary concern is the error in polarization characteristics. Polarization uncertainty provides information correlated with polarization (DOP) reconstruction error, enabling more accurate DOP recovery compared to the other two uncertainty measures, and Table \ref{tab:2} further confirms this observation.

\noindent \textbf{Impact of SD configuration.} We enable the SD model to adapt to the CPDM task while preserving its diffusion prior by injecting LoRA layers into both the VAE and U-Net components of the SD. Table \ref{tab:3} shows several alternative configurations we explored for the SD model. The ``All weight'' setting, in which all parameters are fine-tuned, leads to training failure due to the limited data distribution. When LoRA is injected only into the U-Net, performance degrades, indicating that updating the VAE is also crucial. Therefore, we choose to apply LoRA to both the U-Net and VAE. We also investigate the effect of LoRA rank on performance, when the rank is set to 2, training becomes unstable; when the rank is 4 or higher, performance plateaus. Since larger ranks introduce higher computational and memory overhead, we ultimately set the rank to 4.
\begin{figure}[!t]
	\begin{center}
		\includegraphics[width=\linewidth]{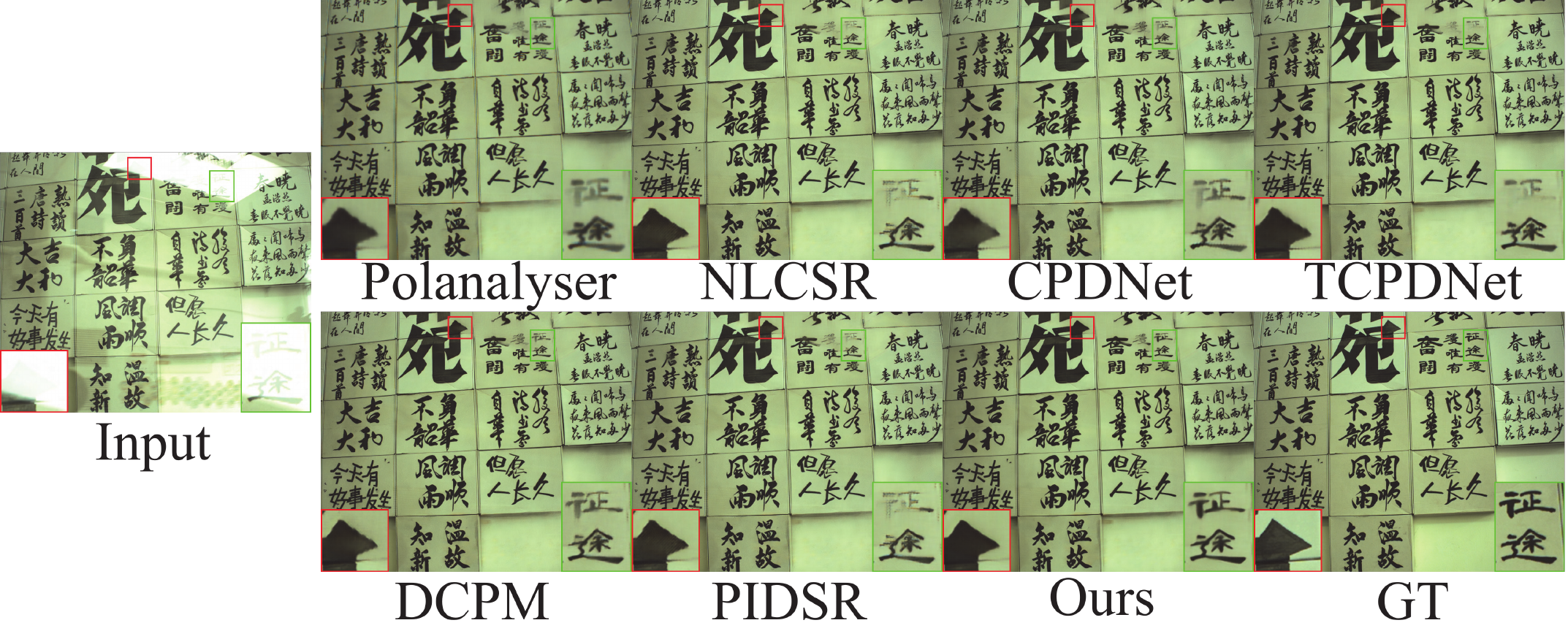}
	\end{center}
	\caption{Visual comparisons for PRR of different methods. Our method achieves the clearest reflection removal results.}
	\label{fig:7}
\end{figure}
\subsection{Applications}
To fully demonstrate the demosaicking performance of our method, we conduct the polarization-based reflection removal (PRR) experiment shown in Fig. \ref{fig:7}. After demosaicking, all images are processed using the method provided by PolarFree \cite{yao2025polarfree} for reflection removal. Our method exhibits the fewest artifacts and the clearest text, demonstrating its strong performance in the PRR task.
\section{Conclusion}
In this paper, we transfer the powerful diffusion prior to the CPDM task via LoRA. Derived from large-scale natural image distributions, the diffusion prior helps overcome the performance bottleneck imposed by the scarcity of demosaicking training data. We propose a dual-branch demosaicking architecture, the outputs of the two branches are adaptively fused through explicitly modelled polarization uncertainty. The uncertainty guides the diffusion branch to refine regions with large polarization errors, while using the base branch in low error regions to maintain fidelity. The proposed method achieves state-of-the-art performance across multiple benchmarks, significantly improving the accuracy of CPDM.

\section*{Acknowledgments}
This work was supported in part by the National Natural Science Foundation of China (Grant No. 62105372),  the National Natural Science Foundation of China (Grant No. 62471497), the Fundamental Research Foundation of National Key Laboratory of Automatic Target Recognition (Grant Number: WDZC20255290209), Hunan Provincial Research and Development Project (Grant No. 2025QK3019). This work was supported in part by the High 
Performance Computing Center of Central South University.

\bibliography{aaai2026}

@inproceedings{zamir2022restormer,
  title={Restormer: Efficient transformer for high-resolution image restoration},
  author={Zamir, Syed Waqas and Arora, Aditya and Khan, Salman and Hayat, Munawar and Khan, Fahad Shahbaz and Yang, Ming-Hsuan},
  booktitle={Proceedings of the Computer Vision and Pattern Recognition Conference},
  pages={5728--5739},
  year={2022}
}

@InProceedings{Hwang_2025_ICCV,
    author    = {Hwang, Inseung and Choi, Kiseok and Ha, Hyunho and Kim, Min H.},
    title     = {Benchmarking Burst Super-Resolution for Polarization Images: Noise Dataset and Analysis},
    booktitle = {Proceedings of the IEEE/CVF International Conference on Computer Vision (ICCV)},
    month     = {October},
    year      = {2025},
    pages     = {24899-24909}
}

@inproceedings{yao2025polarfree,
  title={PolarFree: Polarization-based Reflection-Free Imaging},
  author={Yao, Mingde and Wang, Menglu and Tam, King-Man and Li, Lingen and Xue, Tianfan and Gu, Jinwei},
  booktitle={Proceedings of the Computer Vision and Pattern Recognition Conference},
  pages={10890--10899},
  year={2025}
}

@article{wen2019convolutional,
  title={Convolutional demosaicing network for joint chromatic and polarimetric imagery},
  author={Wen, Sijia and Zheng, Yinqiang and Lu, Feng and Zhao, Qinping},
  journal={Optics Letters},
  volume={44},
  number={22},
  pages={5646--5649},
  year={2019},
  publisher={Optical Society of America}
}

@misc{maeda2019polanalyser,
  author = {Ryota Maeda},
  title = {Polanalyser: Polarization Image Analysis Tool},
  url = {https://github.com/elerac/polanalyser},
  year = {2019},
}

@inproceedings{qiu2021linear,
  title={Linear polarization demosaicking for monochrome and colour polarization focal plane arrays},
  author={Qiu, Simeng and Fu, Qiang and Wang, Congli and Heidrich, Wolfgang},
  booktitle={Computer Graphics Forum},
  volume={40},
  number={6},
  pages={77--89},
  year={2021},
  organization={Wiley Online Library}
}

@book{bowman2012introduction,
  title={Introduction to Bessel functions},
  author={Bowman, Frank},
  year={2012},
  publisher={Courier Corporation}
}

@book{collett2005field,
  title={Field guide to polarization},
  author={Collett, Edward},
  volume={15},
  year={2005},
  publisher={SPIE press Bellingham}
}

@article{chung2022diffusion,
  title={Diffusion posterior sampling for general noisy inverse problems},
  author={Chung, Hyungjin and Kim, Jeongsol and Mccann, Michael T and Klasky, Marc L and Ye, Jong Chul},
  journal={arXiv preprint arXiv:2209.14687},
  year={2022}
}

@article{wang2022zero,
  title={Zero-shot image restoration using denoising diffusion null-space model},
  author={Wang, Yinhuai and Yu, Jiwen and Zhang, Jian},
  journal={arXiv preprint arXiv:2212.00490},
  year={2022}
}

@article{wu2024one,
  title={One-step effective diffusion network for real-world image super-resolution},
  author={Wu, Rongyuan and Sun, Lingchen and Ma, Zhiyuan and Zhang, Lei},
  journal={Advances in Neural Information Processing Systems},
  volume={37},
  pages={92529--92553},
  year={2024}
}

@inproceedings{yue2025arbitrary,
  title={Arbitrary-steps image super-resolution via diffusion inversion},
  author={Yue, Zongsheng and Liao, Kang and Loy, Chen Change},
  booktitle={Proceedings of the Computer Vision and Pattern Recognition Conference},
  pages={23153--23163},
  year={2025}
}

@article{kawar2022denoising,
  title={Denoising diffusion restoration models},
  author={Kawar, Bahjat and Elad, Michael and Ermon, Stefano and Song, Jiaming},
  journal={Advances in Neural Information Processing Systems},
  volume={35},
  pages={23593--23606},
  year={2022}
}

@inproceedings{wu2024seesr,
  title={Seesr: Towards semantics-aware real-world image super-resolution},
  author={Wu, Rongyuan and Yang, Tao and Sun, Lingchen and Zhang, Zhengqiang and Li, Shuai and Zhang, Lei},
  booktitle={Proceedings of the Computer Vision and Pattern Recognition Conference},
  pages={25456--25467},
  year={2024}
}

@article{sun2023improving,
  title={Improving the Stability and Efficiency of Diffusion Models for Content Consistent Super-Resolution},
  author={Sun, Lingchen and Wu, Rongyuan and Liang, Jie and Zhang, Zhengqiang and Yong, Hongwei and Zhang, Lei},
  journal={arXiv preprint arXiv:2401.00877},
  year={2023}
}

@article{chihaoui2024blind,
  title={Blind image restoration via fast diffusion inversion},
  author={Chihaoui, Hamadi and Lemkhenter, Abdelhak and Favaro, Paolo},
  journal={arXiv preprint arXiv:2405.19572},
  year={2024}
}

@inproceedings{sun2025pixel,
  title={Pixel-level and semantic-level adjustable super-resolution: A dual-lora approach},
  author={Sun, Lingchen and Wu, Rongyuan and Ma, Zhiyuan and Liu, Shuaizheng and Yi, Qiaosi and Zhang, Lei},
  booktitle={Proceedings of the Computer Vision and Pattern Recognition Conference},
  pages={2333--2343},
  year={2025}
}

@article{chen2025invertible,
  title={Invertible diffusion models for compressed sensing},
  author={Chen, Bin and Zhang, Zhenyu and Li, Weiqi and Zhao, Chen and Yu, Jiwen and Zhao, Shijie and Chen, Jie and Zhang, Jian},
  journal={IEEE Transactions on Pattern Analysis and Machine Intelligence},
  year={2025},
    volume={47},
  number={5},
  pages={3992-4006},
  publisher={IEEE}
}

@inproceedings{chen2025adversarial,
  title={Adversarial diffusion compression for real-world image super-resolution},
  author={Chen, Bin and Li, Gehui and Wu, Rongyuan and Zhang, Xindong and Chen, Jie and Zhang, Jian and Zhang, Lei},
  booktitle={Proceedings of the Computer Vision and Pattern Recognition Conference},
  pages={28208--28220},
  year={2025}
}

@article{li2019demosaicking,
  title={Demosaicking DoFP images using Newton’s polynomial interpolation and polarization difference model},
  author={Li, Ning and Zhao, Yongqiang and Pan, Quan and Kong, Seong G},
  journal={Optics Express},
  volume={27},
  number={2},
  pages={1376--1391},
  year={2019},
  publisher={Optical Society of America}
}

@inproceedings{morimatsu2020monochrome,
  title={Monochrome and color polarization demosaicking using edge-aware residual interpolation},
  author={Morimatsu, Miki and Monno, Yusuke and Tanaka, Masayuki and Okutomi, Masatoshi},
  booktitle={2020 IEEE International Conference on Image Processing (ICIP)},
  pages={2571--2575},
  year={2020},
  organization={IEEE}
}

@article{luo2023sparse,
  title={Sparse representation-based demosaicking method for joint chromatic and polarimetric imagery},
  author={Luo, Yidong and Zhang, Junchao and Tian, Di},
  journal={Optics and Lasers in Engineering},
  volume={164},
  pages={107526},
  year={2023},
  publisher={Elsevier}
}

@article{luo2024learning,
  title={Learning a non-locally regularized convolutional sparse representation for joint chromatic and polarimetric demosaicking},
  author={Luo, Yidong and Zhang, Junchao and Shao, Jianbo and Tian, Jiandong and Ma, Jiayi},
  journal={IEEE Transactions on Image Processing},
  year={2024},
  volume={33},
  number={},
  pages={5029-5044},
  publisher={IEEE}
}

@article{wen2021sparse,
  title={A sparse representation based joint demosaicing method for single-chip polarized color sensor},
  author={Wen, Sijia and Zheng, Yinqiang and Lu, Feng},
  journal={IEEE Transactions on Image Processing},
  volume={30},
  pages={4171--4182},
  year={2021},
  publisher={IEEE}
}

@article{sun2021color,
  title={Color polarization demosaicking by a convolutional neural network},
  author={Sun, Yuanyuan and Zhang, Junchao and Liang, Rongguang},
  journal={Optics Letters},
  volume={46},
  number={17},
  pages={4338--4341},
  year={2021},
  publisher={Optical Society of America}
}

@article{guo2024attention,
  title={Attention-based progressive discrimination generative adversarial networks for polarimetric image demosaicing},
  author={Guo, Yuxuan and Dai, Xiaobing and Wang, Shaoju and Jin, Guang and Zhang, Xuemin},
  journal={IEEE Transactions on Computational Imaging},
  year={2024},
  volume={10},
  number={},
  pages={713-725},
  publisher={IEEE}
}

@article{li2025demosaicking,
  title={Demosaicking customized diffusion model for snapshot polarization imaging},
  author={Li, Chenggong and Luo, Yidong and Wu, Caiyun and Zhang, Junchao and Yang, Degui and Zhao, Dangjun},
  journal={Optics \& Laser Technology},
  volume={188},
  pages={112868},
  year={2025},
  publisher={Elsevier}
}

@inproceedings{zhou2025pidsr,
  title={PIDSR: Complementary Polarized Image Demosaicing and Super-Resolution},
  author={Zhou, Shuangfan and Zhou, Chu and Lyu, Youwei and Guo, Heng and Ma, Zhanyu and Shi, Boxin and Sato, Imari},
  booktitle={Proceedings of the Computer Vision and Pattern Recognition Conference},
  pages={16081--16090},
  year={2025}
}

@article{zhang2018learning,
  title={Learning a convolutional demosaicing network for microgrid polarimeter imagery},
  author={Zhang, Junchao and Shao, Jianbo and Luo, Haibo and Zhang, Xiangyue and Hui, Bin and Chang, Zheng and Liang, Rongguang},
  journal={Optics Letters},
  volume={43},
  number={18},
  pages={4534--4537},
  year={2018},
  publisher={Optical Society of America}
}

@inproceedings{nguyen2022two,
  title={Two-step color-polarization demosaicking network},
  author={Nguyen, Vy and Tanaka, Masayuki and Monno, Yusuke and Okutomi, Masatoshi},
  booktitle={2022 IEEE International Conference on Image Processing (ICIP)},
  pages={1011--1015},
  year={2022},
  organization={IEEE}
}

@article{luo2025cpifuse,
  title={CPIFuse: Toward realistic color and enhanced textures in color polarization image fusion},
  author={Luo, Yidong and Zhang, Junchao and Li, Chenggong},
  journal={Information Fusion},
  volume={120},
  pages={103111},
  year={2025},
  publisher={Elsevier}
}

@inproceedings{wu2025glossy,
  title={Glossy Object Reconstruction with Cost-effective Polarized Acquisition},
  author={Wu, Bojian and Peng, Yifan and Hu, Ruizhen and Zhou, Xiaowei},
  booktitle={Proceedings of the Computer Vision and Pattern Recognition Conference},
  pages={422--431},
  year={2025}
}

@inproceedings{rebhan2019principle,
  title={Principle investigations on polarization image sensors},
  author={Rebhan, David and Rosenberger, Maik and Notni, Gunther},
  booktitle={Photonics and Education in Measurement Science 2019},
  volume={11144},
  pages={50--54},
  year={2019},
  organization={SPIE}
}

@article{wu2021polarization,
  title={Polarization image demosaicking using polarization channel difference prior},
  author={Wu, Rongyuan and Zhao, Yongqiang and Li, Ning and Kong, Seong G},
  journal={Optics Express},
  volume={29},
  number={14},
  pages={22066--22079},
  year={2021},
  publisher={Optical Society of America}
}

@article{hu2022lora,
  title={Lora: Low-rank adaptation of large language models.},
  author={Hu, Edward J and Shen, Yelong and Wallis, Phillip and Allen-Zhu, Zeyuan and Li, Yuanzhi and Wang, Shean and Wang, Lu and Chen, Weizhu and others},
  journal={ICLR},
  volume={1},
  number={2},
  pages={3},
  year={2022}
}

\end{document}